\documentclass[conference]{IEEEtran}

\usepackage{graphicx}      

\usepackage{siunitx} 
\usepackage{xcolor} 
\usepackage{amssymb,amsfonts,mathtools,tensor,amsmath, amssymb} 
\usepackage{blindtext} 
\usepackage{subfig}

\usepackage{booktabs}
\usepackage{tabu}
\usepackage{multirow}
\usepackage{caption}

\usepackage{pgfplots}
\pgfplotsset{compat=newest}
\usepackage{varwidth}
\usetikzlibrary{arrows}
\usepackage{tikzscale}
\usepackage{xcolor}

\definecolor{mycolor1}{cmyk}{0,0,0,1}%
\definecolor{mycolor2}{cmyk}{1,0,1,0}
\definecolor{mycolor3}{cmyk}{0.2,0.5,0.5,0}
\definecolor{mycolor4}{cmyk}{0,0.2,1,0}
\definecolor{mycolor5}{cmyk}{0.9,0.8,0,0}%
\definecolor{mycolor6}{cmyk}{0.55,0.1,0.9,0}%
\definecolor{mycolor7}{cmyk}{0.3,0.7,0.2,0}%
\definecolor{mycolor8}{cmyk}{0.3,0.5,0.7,0.5}%
\definecolor{green}{cmyk}{0.55,0,0.48,0}%

\pgfplotsset{compat=newest}

\newcommand{\m}[1]{\boldsymbol{#1}} 
\newcommand{\mat}[1]{\begin{bmatrix*}[r]#1 
\end{bmatrix*}}

\renewcommand{\a}{\m{u}} 
\newcommand{\Ai}{U_i} 
\newcommand{\Ac}{U_i} 
\newcommand{\Uphat}{U_{\hat{p}_i}} 
\newcommand{\ascalar}{u} 

\newcommand{\todo}[1]{}
\newcommand{\tododone}[1]{}

\newtheorem{definition}{Definition}

\usepackage{algorithm}
\usepackage{algpseudocode} 
\algnewcommand{\Initialize}[1]{%
	\State \textbf{initialize} {\raggedright #1}
}
\makeatletter
\algnewcommand{\Statey}[1]{\Statex \hskip\ALG@thistlm #1}
\makeatother
\algdef{SE}[DOWHILE]{Do}{doWhile}{\algorithmicdo}[1]{\algorithmicwhile\ #1}%

\def\BibTeX{{\rm B\kern-.05em{\sc i\kern-.025em b}\kern-.08em
		T\kern-.1667em\lower.7ex\hbox{E}\kern-.125emX}}

\title{Partner Approximating Learners (PAL):\\Simulation-Accelerated Learning with Explicit Partner Modeling in Multi-Agent Domains$^{**}$}
	
	\makeatletter
	\newcommand{\linebreakand}{%
	\end{@IEEEauthorhalign}
	\hfill\mbox{}\par
	\mbox{}\hfill\begin{@IEEEauthorhalign}
	}
	\makeatother	
	
	\author{\IEEEauthorblockN{Florian K\"opf$^{*}$\thanks{$^{*}$These authors contributed equally to this work.}, Alexander Nitsch$^{*}$, Michael Flad and S\"oren Hohmann}
		\IEEEauthorblockA{\textit{Institute of Control Systems (IRS)}\\
			\textit{Karlsruhe Institute of Technology (KIT)}\\
			Karlsruhe, Germany \\
			florian.koepf@kit.edu, soeren.hohmann@kit.edu}	
	}
	
\begin{document}
		
\maketitle

{
	\renewcommand{\thefootnote}%
	{\fnsymbol{footnote}}
	\footnotetext[1]{These authors contributed equally to this work.}
}
{
	\renewcommand{\thefootnote}%
	{\fnsymbol{footnote}}
	\footnotetext[7]{This work has been submitted to IEEE for possible publication.}
}

\begin{abstract}                
Mixed cooperative-competitive control scenarios such as human-machine interaction with individual goals of the interacting partners are very challenging for reinforcement learning agents. In order to contribute towards intuitive human-machine collaboration, we focus on problems in the continuous state and control domain where no explicit communication is considered and the agents do not know the others' goals or control laws but only sense their control inputs retrospectively.
Our proposed framework combines a learned partner model based on online data with a reinforcement learning agent that is trained in a simulated environment including the partner model. 
Thus, we overcome drawbacks of independent learners 
and, in addition, benefit from a reduced amount of real world data required for reinforcement learning which is vital in the human-machine context.
We finally analyze an example that demonstrates the merits of our proposed framework which learns fast due to the simulated environment and adapts to the continuously changing partner because of the partner approximation.
\end{abstract}




\section{Introduction}
In numerous control problems such as robotics, intelligent manufacturing plants and highly-automated driving, several so-called agents (e.g. machines and/or humans) are involved and need to adapt to each other in order to improve their behavior. Allowing the agents to pursue individual, not necessarily opposing, goals by means of individual reward structures leads to mixed cooperative-competitive \cite{Lowe.2017} reinforcement learning (RL) problems. Although, especially in control problems, the system dynamics is often known, the reward structures and control laws of other agents are usually unknown to each agent. If the agents adapt their behavior during runtime, this leads to non-stationary environments from the point of view of each agent. Thus, rather than being ignorant concerning the presence of other agents, it is advisable to consider their influence explicitly \cite{Matignon.2012}. 
Another challenge arising when control tasks are learned by means of RL is the lack of data efficiency as much real-world data is required in order to obtain decent performance. Major successes in training an agent in simulated environments rather than solely based on real data have been reported by \cite{Brokaw.2016} and \cite{Andrychowicz.2018}. In simulated environments, powerful hardware can be used to speed up simulations and thus increase the rate of interactions without risk of erratic exploration. However, concerning the multi-agent case, RL based solely on simulations would not appropriately consider the other agents' non-stationary behavior.

In this work, we focus on control problems in continuous state and control spaces where no explicit communication or reward sharing is available but the agents are solely able to sense or deduce the other agents' control inputs after they have been applied. In order to account for the above-mentioned challenges, we propose a general framework that combines the merits of maintaining a partner model that is constantly updated based on real data with learning in a simulated environment. More precisely, each agent approximates a partner model that incorporates the aggregated controls of all other agents. This approximation is constantly updated based on real data in order to capture their changing behavior. Then, each agent simulates a virtual replica of the real control loop containing the system model, the partner approximation and his own control law. In this virtual simulation, the agent updates his control law by means of RL methods in order to improve the performance w.r.t. his reward function. The control law learned in the virtual environment is then transferred to the real control loop and the partner approximation is updated again in order to capture the other agents' changes and reaction. That way, potentially non-stationary partners are steadily approximated and explicitly considered by the RL agent learning in a simulated environment. 

In the next section, we define our problem and the concept of partner approximating learners (PAL). Then, we place our framework in the context of related work before we propose our main topology. Finally, we give an example choice of the components and analyse our method by means of a swing-up task of an inverted pendulum.


\section{Problem and partner approximating learner definition}\label{Section_prob}
Consider a discrete-time system controlled by $N$ agents that is given in nonlinear state space representation $\m{x}_{k+1} = \m{f}\left(\m{x}_k, \a_{1,k},\dots,\a_{N,k}\right)$, where $\m{x}_k\in X\subseteq \mathbb{R}^n$ and $\a_{i,k}\in \Ai\subseteq \mathbb{R}^{q_i}$ are the continuous state and continuous control of agent $i\in\left\{1,\dots,N\right\}$. From the point of view of agent $i$, let $\a_{p_i} = \mat{\a_1^\intercal&\dots&\a_{i-1}^\intercal&\a_{i+1}^\intercal&\dots&\a_{N}^\intercal}^\intercal$ be the aggregated control input of all other agents. Then, agent $i$ aims to adapt his control law $\m{\pi}_i: X \rightarrow \Ai$ in order to maximize his long-term discounted reward
\begin{align}
\begin{aligned}
R_i&=\sum_{k=0}^{\infty}\gamma_i^k r_i\left(\m{x}_k, \a_{i,k},\a_{p_i,k}\right)\\&=\sum_{k=0}^{\infty}\gamma_i^k r_i\left(\m{x}_k, \m{\pi}_{i}(\m{x}_k),\m{\pi}_{p_i}(\m{x}_k)\right).
\end{aligned}
\end{align}
In this mixed cooperative-competitive setting with continuous state and control spaces, no explicit communication between the agents is allowed. Each agent $i$ maintains a model of the system dynamics $\m{f}$,
either as a result of model design or approximated via e.g. recurrent neural networks. Furthermore, each agent senses the partners' controls $\a_{p_i,k-1}$, i.e. after one step delay, as well as the system state $\m{x}_k$ and is aware of his own reward function $r_i\left(\m{x}_k, \a_{i,k},\a_{p_i,k}\right)$ and discount factor $\gamma_i$, but has no knowledge of the other agents' reward functions $r_j(\cdot)$, $j\neq i$ and control laws $\m{\pi}_{p_i}(\m{x}_k)=\a_{p_i,k}$. Based on this problem definition, the notion of partner approximating learners (PAL) is given as follows.

\tododone{Definition PAL überarbeitet. Frage ist: Brauchen wir die Definition? Eigentlich schon ein bisschen, weil wir den Begriff damit einmal klarstellen. Die Frage ist, ob die Definition von PAL hier Sinn macht oder in 4? Für die Abgrenzung zum SdT ist die Definition hier evtl. sinnvoll, da MADDPG dann kein PAL ist und der Vergleich mit Dyna besser gezogen werden kann?}
\begin{definition}\label{de:PAL}
	A \textit{partner approximating learner (PAL}) is an RL agent in a multiagent setting acting in continuous state and control spaces that
	\begin{itemize}
		\item does not explicitly communicate with other agents and does not know their reward structures and control laws
		\item is able to sense or deduce the other agents' control inputs after they have been applied
		\item maintains and updates a model of the other agents' aggregated control law (partner model) based on real data
		\item updates his control law based on simulated data while explicitly incorporating the partner model.
	\end{itemize}
\end{definition}
In the following section, we outline related work before our proposed framework is introduced.


\section{Related work}\label{Section_related_work}

The idea to incorporate simulated data from a system model into the learning process of an RL agent was proposed in the Dyna architecture \cite{Sutton.1991}.
Extensions of this concept to use a simulated environment rather than solely learning from real data marked a breakthrough in order to cope with the sample complexity when using high-dimensional function approximators in continuous control tasks. One example is the use of Normalized Advantage Functions (NAF) with Imagination Rollouts \cite{Gu.2016}, which not only allows for the use of continuous state and control spaces, but accelerates learning by means of model-based simulated data that is additionally fed into the replay buffer.
Another example is given by \cite{Andrychowicz.2018}, where a complex dexterous hand manipulation task has successfully been learned in simulation based on Proximal Policy Optimization (PPO) \cite{Schulman.2017} and transferred to a physical robot.

On the multi-agent side, independent learners have shown limited performance \cite{Matignon.2012} due to the non-stationarity of the environment. In fully cooperative settings, optimistic learning such as hysteretic Q-learning \cite{Matignon.2007} was proposed assuming that all agents tend to improve collective rewards. Other approaches require explicit communication \cite{Foerster.2016, Sukhbaatar.2016} or share actor parameters \cite{Gupta.2017, Hausknecht.2016}.
Partner modeling strives to avoid the disadvantages of independent learners without the necessity of communication or parameter sharing. In Self Other-Modeling \cite{Raileanu.2018}, the agent updates his belief of the partners' hidden goals and predicts the others' controls inputs based on his own control law. In the work of \cite{Foerster.2018}, the maximum-likelihood is used to predict the partners' future controls based on previous controls in finite state and control spaces under the requirement that the payoff matrix is known to all agents.	
Multi-agent Deep Deterministic Policy Gradient MADDPG \cite{Lowe.2017} is a remarkable extension to DDPG \cite{Lillicrap.2015}, thus allowing continuous state and control spaces. MADDPG uses centralized training with decentralized execution. Thus, the Q-function of each agent not only depends on the state and his own control but also the controls of all other agents.
In order to remove the assumption of knowing all agents' control laws, it is suggested in \cite[Section~4.2]{Lowe.2017} to infer control laws of other agents.

Usually, multi-agent RL algorithms do not explicitly assume knowledge of the system dynamics $\m{f}$ and therefore learn only based on observed data. In contrast, our framework benefits from known system dynamics, which is often available in control engineering as a result of model design or can be approximated, and requires real data solely to update the partner model whereas the RL agent is able to explore and generate huge amounts of data in a virtual environment.
Our framework proposed in the next section incorporates partner modeling into the paradigm to accelerate learning by using simulated data and can therefore be interpreted to extend powerful mechanisms such as the Dyna-architecture \cite{Sutton.1991} or Imagination Rollouts \cite{Gu.2016} to the multi-agent case.


\section{Proposed adaptive mixed cooperative-competitive controller} \label{PAL}
\begin{figure}[tb!]             
	\centering 
	\resizebox{\columnwidth}{!}{                  
		\def\svgwidth{300pt}    
		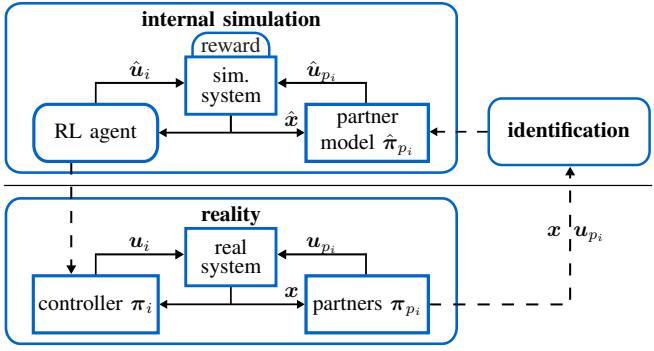}
	\caption{Structure of the proposed framework. Each agent identifies an aggregated partner model from online data, optimizes his control law based on the partner model and system model by means of RL in the internal simulation and transfers the learned control law to the controller in reality.}
	\label{full_structure}
\end{figure}

In this section, we introduce the topology of the proposed Partner Approximating Learner framework (PAL-framework), which can be used with various partner identification and RL algorithms due to its modularity. We refer to all controllers implemented in the PAL-framework as Partner Approximating Learners (PALs).
Our framework consists of three main components that can be seen in Fig.~\ref{full_structure}: the \textit{identification} which approximates all partners' aggregated control law $\m{\pi}_{p_i}$ with a model $\hat{\m{\pi}}_{p_i}$ and the \textit{internal simulation} where RL is used to improve the last component namely the \textit{control law} $\m{\pi}_i$ which is applied to the real physical system to be controlled. In the following, the components will be explained in more detail.


\subsection{Online partner identification with experience replay}\label{identification}

To be able to improve the own control law $\m{\pi}_i$ toward a higher long-term reward $R_i$, the behavior of the partners must be taken into account. 
We therefore continuously identify and improve a model $\hat{\m{\pi}}_{p_i}: X \rightarrow \Uphat$ of $\m{\pi}_{p_i}$ in order to predict the aggregate control input $\a_{p_i,k}$ of all partners from the current state $\m{x}_k$. Note however, that $\m{\pi}_{p_i}$ is not always fixed and might change, e.g. because the partners are learning as well. 
%
Thus, the model $\hat{\m{\pi}}_{p_i}$ should be a flexible and powerful function approximator in order to accurately capture a wide range of possible partner control laws.

Supervised learning algorithms typically require a lot of training data before any useful approximation of the target is obtained. Due to the fact that the data has to be obtained from interactions of the partners with the system, the rate of new information about the partners' behavior is quite low. Additionally, using only the newest set of input-output data $(\m{x}_k, \m{\a}_{p_i,k})$ for training leads to a high variance in the direction of the applied updates to the models which often leads to unstable learning algorithms \cite{Mnih.2013}. 
Both the relative scarcity of data and the high variance of updates also prevented the use of deep neural networks in RL for many years. A breakthrough to both problems was introduced to deep RL by \cite{Mnih.2013} in the form of experience replay (ER). Instead of training on only the latest experience, $m_{\text{RL}}$ samples are chosen from the replay buffer $\mathcal{B}_{\text{RL}}$ uniformly at random (u.a.r.) and form the mini-batch, which is used for training. 

\begin{figure}[tb!]             
	\centering 
	\resizebox{\columnwidth}{!}{                  
		\def\svgwidth{300pt}    
		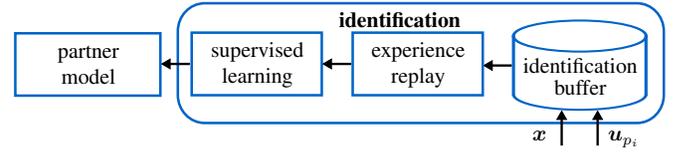}
	\caption{Online identification. Each time step $k$, the state $\m{x}_k$ and control input of the partners $\a_{p_i,k}$ are stored in the identification buffer $\mathcal{B}_{\text{ID}}$. A mini-batch is formed by picking $m_{\text{ID}}$ experiences using an applicable experience replay algorithm and the model $\hat{\m{\pi}}_{p_i}$ of the partners' behavior $\m{\pi}_{p_i}$ is improved.}
	\label{ident_fig}
\end{figure}

Because both online identification and deep RL exhibit these problems, we adapt experience replay for the use in online identification. To this end, we save the input-output data of the partner as experiences $e_k = (\m{x}_k, \a_{p_i,k})$ into an identification buffer $\mathcal{B}_{\text{ID}}$. To update the approximate model of the partner, we pick $m_{\text{ID}}$ experiences from the buffer and use a supervised learning algorithm that is appropriate for the specific task.
The size of the buffer should be large enough to have a high chance of holding information about different regions of the state space and thus capturing nonlinearities in the identification step. Limiting it in size is however not only a memory requirement, but helps to discard experiences that are outdated and thus do not capture the current behavior of the potentially changing partner.
Even improved ER algorithms which differ in the way the experiences are drawn from $\mathcal{B}_{\text{ID}}$, such as prioritized experience replay (PER) \cite{Schaul.2015} and combined experience replay (CER) \cite{Zhang.2017}, can be used directly as long as they do not take the reward of an experience into account (there is no reward associated with the input-output data $e_k = (\m{x}_k, \a_{p_i,k})$ of the partner). To use PER, the priorities are weighted according to the prediction error rather than the TD error.
Fig.~\ref{ident_fig} shows the different components of the identification part of the controller.

\subsection{Internal simulation}
The core idea of the PAL-framework lies within the internal simulation that the controller runs in order to improve its control law. It consists of two parts, a virtual replica of the real control loop and an RL agent acting on this replica.

\subsubsection{Virtual replica}

In order to capture the interactions of the real control loop, the three components of ``reality" in Fig.~\ref{full_structure} have to be known. The controller's behavior $\m{\pi}_i$ and the system dynamics are both known, while the partners' behavior $\m{\pi}_{p_i}$ is not. This is where the approximate partner model $\hat{\m{\pi}}_{p_i}$ (see Section \ref{identification}) is used. We are now able to simulate the behavior of the real control loop offline and typically much faster, with no wear of the hardware and without cumbersome and costly RL on the physical system.

\subsubsection{Reinforcement learning algorithm}
\begin{figure}[tb!]             
	\centering 
	\resizebox{\columnwidth}{!}{                  
		\def\svgwidth{300pt}    
		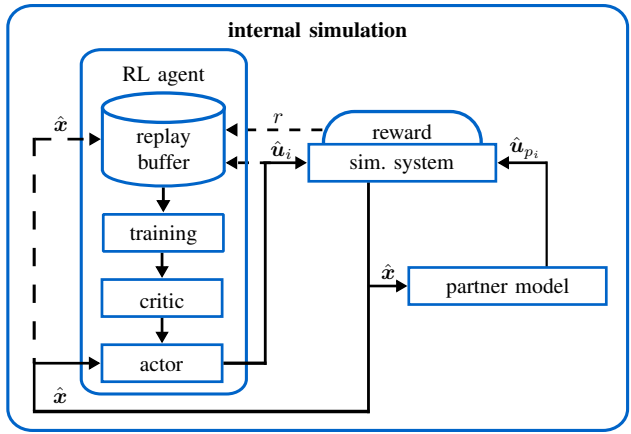}
	\caption{An RL agent improves his control law in the internal simulation based on the partner model and system dynamics.}
	\label{learning_simulation}
\end{figure}

With a simulation of the real control loop at hand, RL can be applied in a straightforward way, when the system and approximate partner model are combined into a single Markov Decision Process (MDP) with state space $X$, action space $\Ac$, system dynamics $\m{f}\left(\hat{\m{x}}_{\hat{k}}, \hat{\a}_{i,\hat{k}},\hat{\pi}_{p_i}(\hat{\m{x}}_{\hat{k}})\right)$, reward  function $r_i\left(\hat{\m{x}}_{\hat{k}},\hat{\a}_{i,\hat{k}},\hat{\pi}_{p_i}(\hat{\m{x}}_{\hat{k}})\right)$ and discount factor $\gamma_i$, where $\hat{k}$ denotes the time step in the simulation.
In this auxiliary MDP, the RL agent chooses simulated controls $\hat{\a}_{i,\hat{k}}$ and obtains the resulting simulated state $\hat{\m{x}}_{\hat{k}}$ of the simulated system. In addition, the agent experiences a reward $r_i\left(\hat{\m{x}}_{\hat{k}},\m{\pi}_i(\hat{\m{x}}_{\hat{k}}),\hat{\m{\pi}}_{p_i}(\hat{\m{x}}_{\hat{k}})\right)$. Based on these experiences, which are usually stored in a replay buffer $\mathcal{B}_{\text{RL}}$, the agent improves his control law. The complete setup of the simulated control loop can be seen in Fig.~\ref{learning_simulation} for the example of an actor-critic RL agent, where the critic estimates the long-term reward and the actor represents the control law. Note that for some RL algorithms, the partner model $\hat{\pi}_{p_i}$ may additionally be used directly by the RL agent, e.g. in the case of MADDPG \cite{Lowe.2017}.

\subsection{Control law $\m{\pi}_i$}
The control law learned in simulation can then be used as the control law of the controller acting on the physical system (i.e. ``reality" in Fig.~\ref{full_structure}).
The representation of the control law $\m{\pi}_i$ that acts on the physical system therefore depends on the kind of RL agent that is used in the internal simulation.
Since the formulation of the problem is done in discrete time, $\m{\pi}_i$ will be used every timestep $k$ to calculate $\a_{i,k}$ for the duration of the next timestep.


\section{Experiments}\label{Section_experiments}
In this section, we give the example system that is used in order to demonstrate the effectiveness of the proposed topology, define concrete algorithms for the experiments and discuss results.
\subsection{Example system} \label{pendulum}
Because of the relevance both in control theory \cite{Astrom.2000} and machine learning \cite{Adam.2012} literature, a pendulum swing-up task is selected.
To easily and reproducibly test the potential of the proposed controller, the ``real”, i.e. physical, system is replaced by a separate simulation, not to be confused with the internal simulation implemented by PALs.
The pendulum has a two-dimensional state space, an angle $\varphi$, where $\varphi = 0=2\pi$ is defined to be the upright position, and an angular velocity $\omega$ and two agents are able to control the pendulum simultaneously.
Both control variables $\ascalar_1$ and $\ascalar_2$ that represent a momentum applied to the pendulum are clipped to the range of $[-5, +5]~ \SI{}{\radian/\second^2}$ which necessitates a swing-up of the pendulum. 
The pendulum model is based on the pendulum from OpenAI Gym \cite{Brockman.2016} and modified to additionally allow a second agent to apply torque to the pendulum. 
At first, the goal is for both controllers to swing-up and hold the pendulum vertically, later we shift the goal to an inclined position.
On reset, the pendulum starts at a random state within $\varphi \in (-\pi, +\pi]~ \SI{}{\radian}$, $\omega \in (-8, +8]~ \SI{}{\radian/\second}$, which means it has some potential and/or kinetic energy at initialization.
The nonlinear system equations are given by
\begin{align} \label{pendulum_equ}
\begin{aligned}
\m{x}_{k+1}=\begin{bmatrix}
\varphi \\
\omega
\end{bmatrix}_{k+1}
&=
\begin{bmatrix}
\varphi + \omega \cdot \Delta t \\
\omega + (\frac{-3g}{2l} \cdot \sin(\varphi+\pi) + \frac{3}{ml^2}) \cdot \Delta t
\end{bmatrix}_k 
\\&\hphantom{=}+
\begin{bmatrix}
0 & 0\\
\Delta t & \Delta t
\end{bmatrix}
\cdot
\begin{bmatrix}
\ascalar_1 \\
\ascalar_2
\end{bmatrix}_k,\nonumber
\end{aligned}
\end{align}

where $\Delta t = \SI{0.05}{s}$, $g = \SI{10}{\metre/\second^2}$, $m = \SI{1}{\kilogram}$ and $l = \SI{1}{m}$.
In the following, concrete algorithms will be chosen to implement PALs.

\subsection{DDPG-PAL}

To approximate the partners, we use a multilayer perceptron (MLP) to be able to capture highly nonlinear control laws $\pi_{p_i}$.
In order to train this partner model $\hat{\pi}_{p_i}$, we use CER \cite{Zhang.2017b}, as it uses new information right when it is available and is fairly robust to the size of the replay buffer.

For the RL agents, the Deep Deterministic Policy Gradient (DDPG) algorithm \cite{Lillicrap.2015} is chosen. This makes the use of continuous state and control spaces possible.
Because of the actor-critic nature of the DDPG algorithm, the control law can also be easily used on the real system, since it is directly available in the form of the actor.
We will refer to this specific PAL implementation as DDPG-PAL.
The choice of optimizers, learning rates and other hyperparameters are given in the supplementary details in Appendix~\ref{supplements}.

\subsection{Examined controller setups}

In order to examine the functionality of DDPG-PALs, the internal simulation, the partner approximation and the RL agent have to work properly. To examine whether all of these components contribute to the proper functioning, several experiments are conducted and presented in the following.
Since we are focusing on interacting agents, both controllers are learning. The metrics that are reported are averaged over ten test runs and the plots are from one of the two runs that were closest to the median. The four different setups that are examined are defined as follows.

\subsubsection{Baseline (no internal simulation; no explicit identification)}
The direct but naive way of using RL for a cooperative swing-up task follows the \textit{independent learner} paradigm (cf. \cite{Matignon.2012}). In this case, both the controller and its partner are regular DDPG agents interacting with the same physical environment without using an internal simulation. In order not to withhold information that the DDPG-PAL possesses, the baseline agents can measure the delayed output of each other and treat it like a third state of the system. 
For the agent, this can reduce the perceived instationarity of the MDP containing an adaptive partner \cite{Lowe.2017}.
\vspace{-0.1cm}
\subsubsection{Oblivious DDPG-agents in a simulated environment (using an internal simulation; identification disabled)}
Since both agents, while initialized differently, have the same goal, it might be possible for them to achieve the swing-up without knowledge of the other controller. To test if the identification is indeed improving the agents' performance, we use the internal simulation while disabling the identification. This results in each controller learning in an internal simulation which only incorporates the system model. They learn as if there were no partner influencing the system, with no way of realizing that there is, which is why we call them oblivious DDPG-agents.
\vspace{-0.1cm}
\subsubsection{DDPG-PALs (using an internal simulation and identification)}
In order to improve both learning time and quality through simulated experience and a partner model, we use DDPG-PAL for both partners. Therefore, this controller setup represents an example of our proposed PAL-architecture.
For the scenarios above, the goal of swinging up the pendulum and holding it upright is expressed with the reward function $r\left(\m{x}_k, \ascalar_{k}\right) = -\varphi_k^2 - 0.1 \omega_k^2 - 0.01 \ascalar_{k}^2$ for both agents, i.e. $\ascalar_{k} \in \left\{\ascalar_{1,k}, \ascalar_{2,k}\right\}$.
It punishes the control effort $\ascalar_{k}^2$, the deviation from the vertical position $\varphi=0=2\pi$ and the angular velocity $\omega$.
\vspace{-0.1cm}
\subsubsection{DDPG-PALs with different reward functions}
While the aforementioned settings serve to examine the advantages of the PAL-architecture, the fourth experiment uses partners with different reward functions. This is motivated by the case of human-robot-collaboration, where, although goals are typically aligned, different humans might prefer different ways of achieving the goal. As an example, imagine the task for a human to transport a piece of equipment from $A$ to $B$ with support by a robot. Understandably, taller people might have different preferences regarding the height it should be transported at compared to shorter people. A suitable robot controller would ideally both realize and account for those preferences and thus learn to support different human partners differently when transporting the piece of equipment as long as this aligns with its own goals.

To mimic this situation, we use two DDPG-PALs with slightly different reward functions. 
Here, the machine controller (agent~1), trying to cooperate with the partner (agent~2), uses the reward
\begin{align}
r_{1}\left(\m{x}_k, \ascalar_{1,k}\right) = -\left(\left|\varphi_k \right| - \frac{\pi}{4}\right)^2 - 0.1 \omega_k^2 - 0.1 \ascalar_{1,k}^2.
\end{align}
Thus, agent~1 has two optima for the pendulum position, $\varphi_{\text{opt}}$ and $-\varphi_{\text{opt}}$. Note that $\varphi_{\text{opt}} \approx 0.3$ is the angle at which the negative reward caused by the deviation from $\pi / 4$ and the constant control effort to hold this position are balanced.
On the other hand, the partner (e.g. representing the human) uses
\begin{align}
r_{2}\left(\m{x}_k, \ascalar_{2,k}\right) = -\left(\varphi_k - \frac{\pi}{4}\right)^2 - 0.1 \omega_k^2 - 0.1 \ascalar_{2,k}^2.
\end{align}
This means he would like to swing-up the pendulum and hold it at $\varphi_{\text{opt}}$. The second optimum for agent~1 at $-\varphi_{\text{opt}}$ thus leads to a lower reward for agent~2.  To ease the swing-up for this task, the limits of $\ascalar_1$ and $\ascalar_{2}$ are widened to $\left[ -10, 10\right] \SI{}{\radian/{\second^2}}$.

\subsection{Results}
\begin{figure}[tb!]
	\begin{center}	
		\includegraphics[width=\columnwidth, height=5cm]{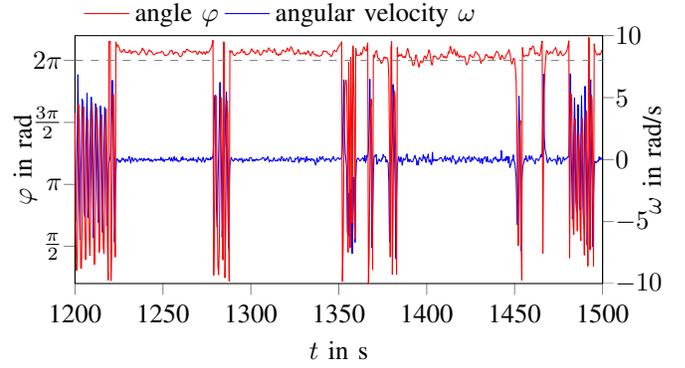}
		\caption{Cooperative swing-up with two baseline agents. Note the shifted time axis.}\label{reality_learner_plot}
	\end{center}
\end{figure}

As is depicted in Fig.~\ref{reality_learner_plot}, baseline DDPG-agents swing-up the pendulum towards the vertical position $\varphi = 0=2\pi$ at around $\SI{1220}{s}$ for the first time. In addition to taking relatively long until the first successful swing-up is performed, holding the pendulum upright is very unstable and it can be seen that the pendulum tips over multiple times with no significant improvement.

\begin{figure}[tb!]
	\begin{center}	
		\includegraphics[width=\columnwidth, height=5cm]{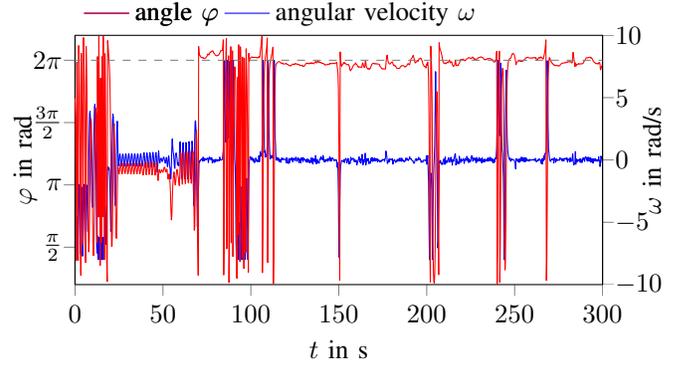}
		\caption{Two oblivious DDPG-agents performing the swing-up task without knowledge of each other.}\label{oblivious_pals}
	\end{center}
\end{figure}

Considering the \textit{oblivious DDPG-agents}, Fig.~\ref{oblivious_pals} shows that even without the identification a swing-up can generally be learned much faster in the simulated environment compared to the baseline. However, because the impact of the partner is ignored, the pendulum can not be held upright for longer periods of time. This leads to an average reward per second of $-61.8$ over all runs in the first $\SI{300}{\second}$ whereas for the baseline the average reward in this time is $-1104$ and clearly much worse as the pendulum is not held in the upright position at all during this time.

\begin{figure}[tb!]
	\begin{center}	
		\includegraphics[width=\columnwidth, height=5cm]{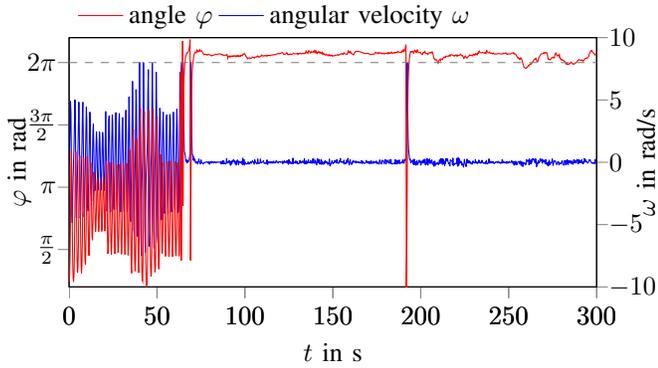}
		\caption{Both agents learn using DDPG-PAL including internal simulations.}\label{both_learn_plot}
	\end{center}
\end{figure}

Fig.~\ref{both_learn_plot} reveals that the swing-up is successful after just $\SI{70}{s}$ when using \textit{DDPG-PALs}. In addition, it can be seen that the pendulum is held upright more stable compared to the baseline and the oblivious DDPG-agents and is easily re-erected after tipping over.
The cooperating DDPG-PALs achieve an average reward per second of $-36,\!79$ which is a significant improvement compared to the oblivious DDGP-agents which do not include the partner model.
\begin{figure}[tb!]
	\begin{center}	
		\includegraphics[width=\columnwidth, height=5cm]{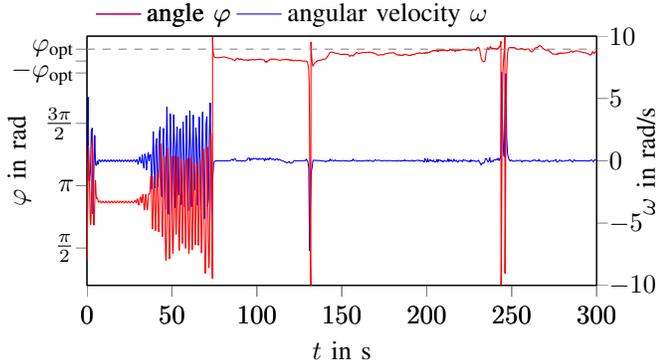}
		\caption{Two DDPG-PALs with different goals agree on the optimum that suits both.}\label{negotiating_pal}
	\end{center}
\end{figure}
These results show that not only the internal simulation, but also the identification significantly improves the results.

For \textit{DDPG-PALs with different reward functions}, Fig.~\ref{negotiating_pal} shows that swinging up the pendulum is also achieved quite fast. At $t = \SI{75}{s}$, the pendulum is held vertically, which already leads to fairly high reward. This vertical position is, however, not the optimum for either of the controllers. Right after tipping over at roughly $t = \SI{130}{s}$, they agree on the optimal position at around $\varphi_{\text{opt}}$.

\subsection{Discussion}
The results above indicate that the desired behavior can successfully be learned by PALs. In order to make broader claims about the applicability especially in the case of PALs with different reward functions, it is necessary to show that agent~1 has indeed learned to prefer $\varphi_{\text{opt}}$ over $-\varphi_{\text{opt}}$, even though this does not follow directly from $r_1$.
Instead, preferring $\varphi_{\text{opt}}$ is better for agent~1 because agent~2 is uncooperative at $-\varphi_{\text{opt}}$, which leads to a lower reward for agent~1 in the region around $-\varphi_{\text{opt}}$.
The preferences of agent~1 can not only be found by experimentation, but explicitly in the DDPG critic, i.e. the action-value function $Q_1\left(\m{x}, \a_1\right)$, that is used in the internal simulation. 
Removing the dependency on the control (i.e. action), we get the state-value function $V_1\left(\m{x}\right)= \max \limits_{\a_1} \left(Q_1\left(\m{x}, \a_1\right)\right)$, which allows us to compare the value that the agent assigns to the two states at $\pm \varphi_{\text{opt}}$ (with $\dot{\varphi}_{\text{opt}} = 0$).
As reference, we perform ten runs where agent~1 has his partner approximation disabled, i.e. follows the oblivious-agent mechanism. This leads to the average values of $V_1\left(\m{x} = (-\varphi_{\text{opt}}, 0) \right) = -35.52$ and $V_1\left(\m{x} = (+\varphi_{\text{opt}}, 0) \right) = -34.69$

which makes a difference of only $0.83$, meaning that the controller does not significantly prefer one of the states over the other. Furthermore, note that this value is solely based on the estimation of the agent in the internal simulation and not on actual rewards. With partner approximation disabled, the MDP in the internal simulation is less complex because the influence of the partner is missing. This leads to the agent estimating higher rewards than he would actually get when acting in the real world where the partner influences the system as well (cf. the oblivious DDPG-agents that suffer from the lack of an appropriate partner model).

When using the full DDPG-PAL algorithm with partner approximation, the distinction becomes much more significant as $V_{1,\text{PAL}}\left(\m{x} = (-\varphi_{\text{opt}}, 0) \right) = -47.55$ and $V_{1,\text{PAL}}\left(\m{x} = (+\varphi_{\text{opt}}, 0) \right) = -39.83$ and reflects reality much better where $-\varphi_{\text{opt}}$ is penalized.
Thus, the agent developed understanding of the situation. This indicates that PALs can indeed learn the preferences of their partner and subsequently improve the control law towards goal-oriented cooperation.

\section{Conclusion}\label{Section_conclusion}
This work introduces a framework named Partner Approximating Learners (PAL-framework) which combines learning the partners' behavior in mixed cooperative-competitive settings under restricted information with deep RL in a simulated environment. The framework offers two major benefits over independent learners merely training on online data. On one hand, maintaining and constantly updating an explicit model of the partners' aggregated control law takes their influence into consideration and allows the agents to adapt to each other. On the other hand, PALs learn in a simulated environment where the current partner model is explicitly used. Thus, PALs reduce wear on the system explore the state space more safely, while relying on the latest partner model. After proposing our framework, we show its merits by an example of a pendulum swing-up task. Here, utilizing the simulated environment rather than simply working on online data significantly speeds up learning. Furthermore, maintaining a partner model improves the performance. Finally, two DDPG-PALs, where one is indifferent to two states and the other prefers one state over another, successfully assess the situation and agree on a reasonable solution despite the challenging setting of the agents having different reward functions.


\bibliographystyle{ieeetr}
\bibliography{citavi3}             
\newpage                                              
\appendix
\section{Supplementary Details}\label{supplements}
The hyperparameters of the identification algorithm as well as the RL agent are given below.
\begin{table}[h!]\captionsetup{width=\columnwidth}\caption{Hyperparameters of the identification.}\label{tab:ident}
	\begin{center}		
		\begin{tabular}{ll}
			\hline
			hyperparameter & value \\ \hline
			time steps between ident. updates &	1 	\\
			learning episodes per ident. update & 	4\\
			number of hidden layers & 3 \\
			neurons per hidden layer & 	16 	\\
			size of identification buffer $\mathcal{B}_{\text{ID}}$ & last $\SI{100}{\second}$ of ``reality"\\
			experience replay & CER \cite{Zhang.2017} 	\\
			training data per ident. update & 	$10\%$ of buffer 	\\
			mini batch size & 20 \\
			initial weights hidden layer & 	u.a.r. $\in \left[-1,1 \right]$ \\
			initial weights output layer  & u.a.r. $\in \left[-10^{-4}, 10^{-4}\right]$ \\
			activation function hidden layer  	& 	sigmoid \\
			activation function output layer   	&  	linear \\
			optimizer  	&  	Adam, learning rate $0.01$, \\&$\beta_1=0.9$, $\beta_2=0.999$, no \\&gradient clipping, decay, \\&fuzz factor or AMSGrad\\
			error metric  & MSE\\
			size of validation set 	&  	0 \\
			shuffle mini batch before training   &  	true  \\ \hline
		\end{tabular}
	\end{center}
\end{table}

\begin{table}[h!]\captionsetup{width=\columnwidth}\caption{Hyperparameters of the DDPG agents. Here, A/C stands for ``actor and critic"}\label{tab:RL}
	\begin{center}		
		\begin{tabular}{ll}
			\hline
			hyperparameter & value \\ \hline
			time steps between RL updates &	2 	\\
				size of replay buffer $\mathcal{B}_{\text{RL}}$ & 	last $\SI{10}{\second}$ of ``reality"\\
				length episode RL training 	& 	$\SI{10}{\second}$ simulated time,\\&$m_{\text{RL}} = 200$\\
				number of hidden layers A/C 	& 	3 \\
				neurons per hidden layer actor   	& 	16 	\\
				neurons per hidden layer critic & 	32 		\\
				activation f. hidden layer A/C & 	sigmoid \\
				activation f. output layer A/C  	& 	linear 	\\
				initial weights all layers A/C   & 	u.a.r. $\in \left[-1,1 \right]$ \\
				optimizer A/C 	&  	Adam, learning rate $0.001$, \\&$\beta_1=0.9$, $\beta_2=0.999$ \\ & gradient clipping $=1.0$\\ & no decay, fuzz factor\\&or AMSGrad \\
				experience replay   	& 	u.a.r. \cite{Lillicrap.2015} 	\\
				discount factor $\gamma$  	& 	$0.99$ 	\\
				batch size     	& 	$32$ \\
				warm up A/C 	& 	100 \\	
				error metric &  MAE	\\
				target network update rate  	&  	$0.001$ \\
				exploration &  	Ornstein-Uhlenbeck with
				\\ & $\theta = 0.15$, $\mu = 0$, $\sigma = 0.3$  \\ \hline
		\end{tabular}
	\end{center}
\end{table}

\end{document}